\documentclass[aps,twocolumn,prl,superscriptaddress,showpacs]{revtex4-1}
\usepackage[T1]{fontenc}
\usepackage[latin9]{inputenc}
\usepackage{times}
\usepackage{color}
\usepackage{xspace}
\usepackage{amssymb,amsmath}
\usepackage{amsbsy}
\usepackage{graphicx}
\usepackage{multirow}
\usepackage{rotating}
\usepackage{microtype}
\usepackage{tabularx}
\usepackage{epstopdf}
\usepackage{float}
\usepackage{mathrsfs}
\usepackage{hyperref}
\usepackage[normalem]{ulem}
\def\be{\begin{equation}}
\def\ee{\end{equation}}
\def\e#1{\label{#1}\end{equation}}
\def\bea{\begin{eqnarray}}
\def\eea{\end{eqnarray}}

\begin{document}
\title{Spin-phonon interfaces in coupled nanomechanical cantilevers}

\author{Thomas Oeckinghaus}
\affiliation{3. Physikalisches Institut, University of Stuttgart, Stuttgart, Germany}
\author{S. Ali Momenzadeh}
\affiliation{3. Physikalisches Institut, University of Stuttgart, Stuttgart, Germany}
\author{Philipp Scheiger}
\affiliation{3. Physikalisches Institut, University of Stuttgart, Stuttgart, Germany}
\author{Tetyana Shalomayeva}
\affiliation{3. Physikalisches Institut, University of Stuttgart, Stuttgart, Germany}
\author{Amit Finkler}
\affiliation{3. Physikalisches Institut, University of Stuttgart, Stuttgart, Germany}
\affiliation{Department of Chemical and Biological Physics, Weizmann Institute of Science, Israel}
\author{Durga Dasari}
\affiliation{3. Physikalisches Institut, University of Stuttgart, Stuttgart, Germany}
\affiliation{Max Planck Institute for
	Solid State Research, Stuttgart, Germany}
\author{Rainer St\"ohr}
\affiliation{3. Physikalisches Institut, University of Stuttgart, Stuttgart, Germany}
\affiliation{Center for Applied Quantum Technology, University of Stuttgart, Stuttgart, Germany}
\author{J\"org Wrachtrup}
\affiliation{3. Physikalisches Institut, University of Stuttgart, Stuttgart, Germany}
\affiliation{Max Planck Institute for
	Solid State Research, Stuttgart, Germany}

\date{\today}

\begin{abstract}
Coupled micro- and nanomechanical oscillators are of fundamental and technical interest for emerging quantum technologies. Upon interfacing with long-lived solid-state spins, the coherent manipulation of the quantum hybrid system becomes possible even at ambient conditions. While, the ability of these systems to act as a quantum bus inducing long-range spin-spin interactions has been known, the possibility to coherently couple electron/nuclear spins to the common modes of multiple oscillators and map their mechanical motion to spin-polarization has not been experimentally demonstrated. We here report experiments on interfacing spins to the common modes of a coupled cantilever system, and show their correlation by translating ultra-low forces induced by radiation from one oscillator to a distant spin. Further, we analyze the coherent spin-spin coupling induced by the common modes and estimate the entanglement generation among distant spins.
\end{abstract}
\maketitle


Mechanical resonators shaped as cantilevers, drums and doubly clamped beams at micro and nanoscale have shown unparalleled applications both as high sensitive mass detectors~\cite{chaste_nanomechanical_2012,hanay_single-protein_2012,malvar_mass_2016,sage_single-particle_2018, ali_momenzadeh_thin_2016, barson_nanomechanical_2017} and as ideal systems for cavity QED experiments~\cite{aspelmeyer_cavity_2014}. When coupled to light these resonators get sensitive to the radiation pressure allowing them to get cooled to the quantum regime with only few phonons~\cite{chan_laser_2011,teufel_sideband_2011,rocheleau_preparation_2010,riviere_optomechanical_2011,groblacher_demonstration_2009,park_resolved-sideband_2009,schliesser_resolved-sideband_2009,verhagen_quantum-coherent_2012,delic_cavity_2019,hong_hanbury_2017}. Using such cooled micromechanical motion of the cantilevers one could generate photon-phonon, phonon-phonon and photon-photon entanglement~\cite{lee_entangling_2011,palomaki_entangling_2013,riedinger_non-classical_2016,marinkovic_optomechanical_2018,riedinger_remote_2018}. Spin-mechanical systems are analogous to optomechanical systems merging magnetism and mechanical motion. Here mechanical objects (modes) are detected either by magnetic gradient force interactions, or strains induced through magnetostrictive effects \cite{hartmann_magnetic_1999, quandt_magnetostrictive_2000}. While coupling between single spins and mechanical motion has been achieved \cite{wilson-rae_laser_2004,balasubramanian_nanoscale_2008,arcizet_single_2011, kolkowitz_coherent_2012} the ability to extend this coupling to multiple spins \cite{rabl_quantum_2010}, so as to achieve phonon mediated coupling (entanglement) among spins has become a far reaching goal. The key limiting factor has been the spin-phonon coupling strength quantified through cooperativity \cite{lee_topical_2017}. With improved fabrication methods, and operational conditions such as low temperature and ultra-high vacuum, the cooperativity could be made larger than one, allowing for coherent spin-phonon dynamics, but the possibility to scale the architecture to multiple cantilevers and couple spins to their common modes has not been explored. While there are proposals to estimate the phonon mediated scaling of spin networks and their possible application for multi-partite entanglement and quantum simulators \cite{rabl_quantum_2010, bennett_phonon-induced_2013}, demonstrating the basic element of such architecture is still missing, i.e, coupling two spins in individual cantilevers to the common modes of a double cantilever system. Here we address this challenge, by studying a coupled cantilever system and show the coupling of spins to the common modes of the cantilevers. 

Over the past decade spin defects in diamond have emerged as ultra-sensitive magnetometers achieving up to a chemical-shift resolution in ambient conditions~\cite{doherty_nitrogen-vacancy_2013, taylor_high-sensitivity_2008,balasubramanian_ultralong_2009,aslam_nanoscale_2017}. Further, the remarkable mechanical properties make diamond a pristine material for high-quality-factor micro- and nanomechanical systems~\cite{ovartchaiyapong_high_2012,tao_single-crystal_2014,rath_diamond_2015}. Diamond mechanical resonators have been successfully fabricated in the shape of singly and doubly clamped nanobeams, bulk resonators, phononic crystals and nanodisks. Coupling of spins associated with nitrogen-vacancy centers to mechanical modes is well established. NVs can e.g. be coupled to the mechanical degrees of freedom of cantilevers using strong magnetic field gradients~\cite{arcizet_single_2011,kolkowitz_coherent_2012}. Alternatively, a spin-mechanical coupling can be established via the intrinsic strain accompanying a deformation of the mechanical structure. Using this approach NV spins were successfully coupled to cantilevers~\cite{ovartchaiyapong_dynamic_2014,lee_strain_2016,meesala_enhanced_2016,teissier_strain_2014,barfuss_strong_2015}, bulk resonators~\cite{macquarrie_mechanical_2013,r._macquarrie_coherent_2014,macquarrie_continuous_2015} and surface acoustic waves~\cite{golter_coupling_2016,golter_optomechanical_2016}. It was shown that this coupling mechanism can be used to drive dipole forbidden spin state transitions and increase the NV spin coherence time by continuous dynamical decoupling~\cite{macquarrie_mechanical_2013,r._macquarrie_coherent_2014,macquarrie_continuous_2015,barfuss_strong_2015}. Furthermore, advanced applications like phonon-cooling and -lasing, quantum networks, spin squeezing and quantum-enhanced sensing have been theoretically studied~\cite{rabl_cooling_2010,kepesidis_phonon_2013,habraken_continuous_2012,wilson-rae_laser_2004,rabl_quantum_2010,schuetz_universal_2015,barson_nanomechanical_2017}. It was also shown that liberational modes of levitating magnetic particles could be coupled to nearby diamond spins, finding applications in ultra-sensitive gyroscopy~\cite{huillery_spin-mechanics_2019}.

\begin{figure*}
\includegraphics[width=1.0\textwidth]{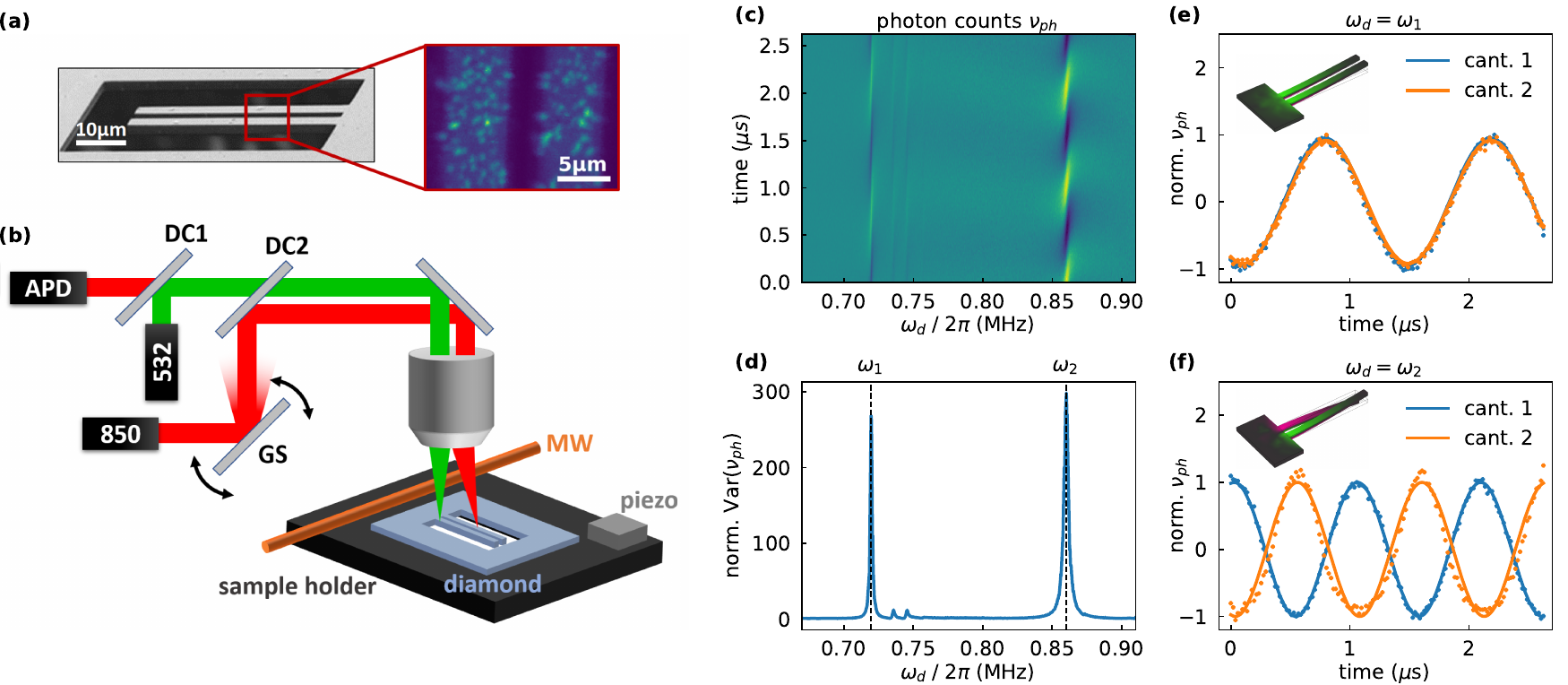}
	\caption{
	(a) SEM image of a typical cantilever pair investigated in this work. The zoom in shows a confocal image of the NVs inside the cantilevers. (b) Sketch of the experimental setup. The photo-thermal driving laser (red) is coupled into the optical path of the readout laser (green) by dichroic mirror (DC2). Using a galvo-scanner (GS), the position of the drive laser beam on the sample surface can be changed relative to the readout laser beam. This enables excitation and readout at two different positions. (c) Time-correlated photon count $\nu_{ph}$ for different driving frequencies $\omega_d$. The photon count rate is proportional to the cantilever deflection. Thus, resonances of the cantilever pair can be seen in the optical signal. (d) Mechanical spectrum of a cantilever pair generated by calculating the variance of the measured photon counts. Two prominent resonances at $\omega_1 = 2\pi \times 0.72$ MHz and $\omega_2 = 2\pi \times 0.86$ MHz are visible. (e) Photon histograms subsequently measured on both cantilevers at the first resonance ($\omega_d=\omega_1$). The cantilevers oscillate in phase (f) Same measurement as shown in (e) but at the second resonance ($\omega_d = \omega_2$). The cantilevers oscillate with a phase shift of $\pi$.
}
\label{fig:fig1}
\end{figure*}

An array of coupled mechanical resonators, each coupled to a separate spin, is a potential layout for future quantum networks. In this case, the spin-spin interaction is mediated by the common modes of the mechanical structure. Such a system could e.g. be implemented as an array of $N$ elastically coupled cantilevers, where the coupling is established by a thin shared base plate holding the cantilevers. 
In this work, we experimentally study the smallest building block of such a network - a two cantilever array $(N=2)$. This system can be described as two harmonic oscillators with spring constant $k$ and coupling constant $k'$. The combined system has two common modes: (i) the symmetric mode where both cantilevers oscillate in-phase at frequency $\omega_1=\sqrt{k/m}$ and (ii) the anti-symmetric mode, where the cantilevers oscillate out of phase at frequency $\omega_2=\sqrt{(k+2k')/m}$.

The cantilevers are held by a $\sim 4\mu$m thin diamond membrane as shown in Fig \ref{fig:fig1}a. The sample hosts a random distribution of shallow nitrogen-vacancy centers (NVs) created by ion-implantation. The opposite side is coated with a 200nm layer of titanium, enabling photo-thermal driving of the cantilevers. To optically excite oscillations, a pulsed IR diode laser (850nm) with a duty cycle of 0.5 is focused onto one of the cantilevers. A galvo scanner allows to move the IR laser beam relative to the beam of the 532nm readout laser (see Fig \ref{fig:fig1}b). Absorption of the IR photons results in a local heating of the cantilever. Due to the different expansion coefficients of the two layers, this leads to a bending of the cantilever (see supplementary information). Alternatively, the system can be excited using a piezo actuator attached to the sample holder.

We first readout the mechanical motion of the cantilevers optically using a 532nm laser. Photons reflected by the sample surface are collected by an APD and time-correlated to the mechanical drive signal. By positioning the laser focus slightly above the sample surface, the photon count rate can be made proportional to the deflection of the cantilever over a range of \textasciitilde300nm. This enables a quantitative readout of the mechanical motion (see supplementary information). From the acquired photon histograms the amplitude, frequency and relative phase of the cantilever oscillation are determined. To obtain the spectrum, we repeat the measurement at different excitation frequencies $\omega_d$ of the photo-thermal driving. Fig \ref{fig:fig1}c shows a photon count map acquired from a cantilever pair. By calculating the variance of the count rate for each frequency, a spectrum is retrieved (see Fig \ref{fig:fig1}d). From the spectrum, we determine the common mode frequencies, $\omega_1=2\pi\times0.72$ MHz and $\omega_2=2\pi\times0.86$ MHz, as well as the quality factor $Q=\omega_2/\Delta\omega\approx 480$. The symmetric and anti-symmetric oscillations of the two cantilevers when driven resonantly at these frequencies are shown in Fig \ref{fig:fig1}e, Fig \ref{fig:fig1}f. In a low temperature (4.8 K) and ultra-high vacuum ($2\times10^{-9}$ mbar) environment, the quality factor increases by a factor of \textasciitilde$10^3$ to $Q_{LT,UHV}\approx 0.4\times10^6$ (see supplementary information).

\begin{figure*}
    \includegraphics[width=1.0\textwidth]{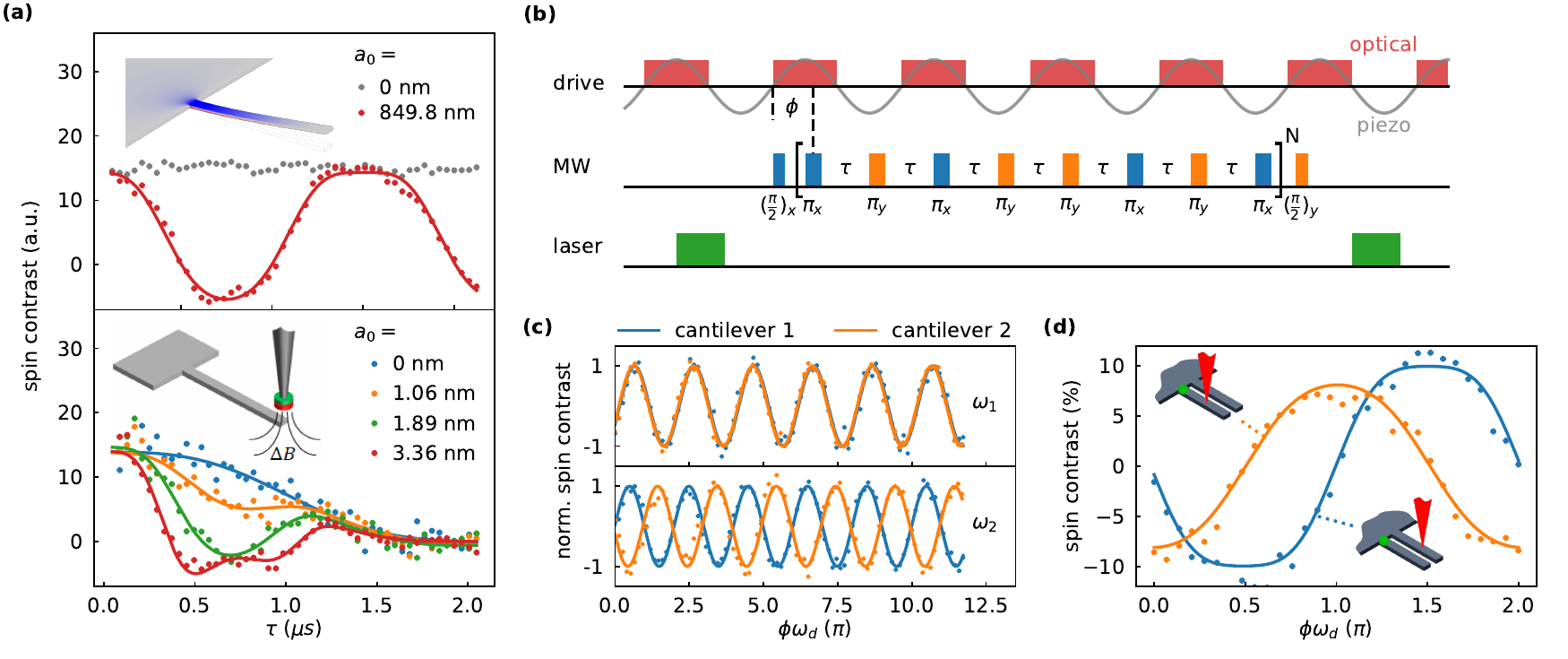}
    \caption{
    (a) Hahn echo measurement of a NV coupled to the mechanical motion via strain (upper) or a magnetic field gradient produced by a magnetized AFM tip (lower). For each coupling mechanism, the signal arising from different oscillation amplitudes were measured. By fitting, the spin-phonon couplings $g_{str} = 2\pi \times 3.0 \pm 0.1$ mHz and $g_{mag} = 7.7 \pm 0.5$ Hz can by determined. (b) Phase-locked dynamical decoupling sequence. The XY8-N dynamical decoupling sequence has a fixed delay $\phi$ to the mechanical drive signal. The central frequency of the decoupling filter function is tuned to the drive frequency $\omega_d$ by fixing the $\pi$-pulse spacing $\tau=\frac{2\pi}{\omega_d}$. At the end of the XY8-N sequence the accumulated phase $\theta$ is projected into the spin polarization by a $(\frac{\pi}{2})_y$ pulse. As a result, the sign of $\theta$ is imprinted into the sign of the spin polarization. (c) Imprint of mechanical oscillations into the spin polarization of a NV in each cantilever using a phase-locked XY8-1 sequence. The oscillations arise from a resonant excitation of the symmetric (upper) and anti-symmetric (lower) mode with the piezo-actuator. (d) Spin contrast of a NV in cantilever 1 acquired with a phase-locked XY8-5 sequence. The mechanical system is excited by photo-thermal driving of cantilever 1 (orange) or cantilever 2 (blue). The results demonstrate the phonon-mediated interaction between a local force acting on cantilever 2 with a distant spin in cantilever 1.
}
    \label{fig:fig2}
\end{figure*}

In the next step we demonstrate the coupling of a NV spin to the mechanical degrees of freedom. There are two well-established methods to achieve this spin-mechanical coupling - strain-induced coupling and magnetic gradient field coupling~\cite{lee_topical_2017}.
The deformation of the diamond lattice caused by a deflection of the cantilever applies axial ($\epsilon_\parallel$) and transverse ($\epsilon_\perp$) strain to embedded NVs~\cite{ovartchaiyapong_dynamic_2014,teissier_strain_2014,barfuss_strong_2015}. This results in a coupling between the NV electron spin and the oscillation of the cantilever. To suppress effects of transverse strain we applied an external magnetic field of \textasciitilde100 G along the NV axis. The strain produces an effective electric field at the NV position resulting in the strain-coupling Hamiltonian 
$H_{str}=(D_0+d_\parallel \epsilon_\parallel ) S_z^2$.
Here, $D_0$ is the NV zero-field splitting, $d_\parallel$ is the parallel strain-coupling constant and $\epsilon_\parallel$ is the strain along the NV axis. 

Fig\ref{fig:fig2}a (upper) shows the imprint of a mechanical oscillation with amplitude $a_0$ into the spin signal produced by a Hahn echo sequence with varying waiting time $\tau$. The system was excited using the piezo-actuator. The oscillation amplitude was determined to be 849.8 nm using the previously introduced quantitative optical readout. From the fitted curves the energy shift $dE = d_\parallel \epsilon_\parallel$ experienced by the NV spin for each oscillation amplitude can be extracted~\cite{bennett_measuring_2012} (see supplementary information). In the small damping regime, the strain is proportional to the cantilever deflection. Thus, the ratio $g_{str}'= dE/a_0 = 2\pi \times (1.21 \pm 0.05)$ kHz/nm is constant and links the mechanical oscillation amplitude to the energy shift of the NV levels. With the zero-point fluctuation of the mechanical system $x_{ZPF}=\sqrt{\hbar/(2m_{eff}\omega)} \approx 1.77\times 10^{-15}$ m results the spin-phonon coupling $g_{str} = g_{str}' x_{ZPF}= 2\pi \times (2.14 \pm 0.08)$ mHz.

The spin-phonon coupling can be increased by creating a strong magnetic field gradient $\delta B_z$ at the apex of the cantilever. The mechanical motion $z(t)$ of an embedded NV is then translated into an effective ac magnetic field $B_{AC} (t)=\delta B_z z(t)$ at the NV's position. Neglecting static magnetic fields, the interaction Hamiltonian is $H_{mag} = g_{mag} (a + a^\dagger) S_z$ with the magnetically-induced spin-phonon coupling $g_{mag} = \gamma_{NV} \delta B_z x_{ZPF}$. In our experiment, a small magnetized tip was placed \textasciitilde50nm above a NV center using an atomic force microscope. Fig\ref{fig:fig2}b (lower) shows Hahn echo measurements for different oscillation amplitudes. Analogous to the strain-induced coupling, $g_{mag}' = \gamma_{NV} \delta B_z = 2\pi \times (0.49 \pm 0.03)$ MHz/nm and $g_{mag} = 2\pi \times (0.87 \pm 0.03)$ Hz can be determined from the fitted curves. Furthermore, the magnetic field gradient $\delta B_z = (1.8 \pm 0.1) \times 10^5$ T/m is extracted. The magnetic field gradient enhances the spin-phonon coupling by a factor of $g_{mag}/g_{str} \approx 2000$. This ratio is specific to the geometry of the cantilevers and the exact positions and orientations of the used NVs. We note that the coherence time decreased by \textasciitilde50\% due to magnetic noise originating from the magnetic tip.

To demonstrate that the NV spin is indeed coupled to a common mode, we imprint the oscillation phase into the spin signal. For this purpose, we employ a phase-locked XY8-N sequence. The used measurement sequence is depicted in Fig \ref{fig:fig2}b. First a $(\pi/2)_x$-pulse creates a superposition state $|\Psi_i\rangle=\frac{1}{\sqrt{2}} (|0\rangle+|-1\rangle)$. Afterwards, the state evolves in between refocusing pulses with a pulse time-spacing $\tau$ into a final state $|\psi_f\rangle=\frac{1}{\sqrt{2}} (|0\rangle+e^{i\phi} |-1\rangle)$. Finally, the phase $\phi$ accumulated by the spin is projected onto the spin polarization by a final $(\pi/2)_y$-pulse.

In the following measurements the inter-pulse spacing $\tau$ of the XY8 sequence is set to match half the period of the cantilever drive $\tau=\pi/\omega_{d}$. Thus, the filter function of the decoupling sequence is always tuned into the cantilever frequency $\omega_{d}$.
By varying the delay $\phi$ between the mechanical drive and the measurement sequence, the motion of the cantilevers can be observed via the spin-mechanical coupling. Fig \ref{fig:fig2}c shows the NV spin signal for varying $\phi$ for a NV in each cantilever, while the system is driven resonantly at the symmetric common mode using the piezoelectric actuator. The measurement shows an in-phase oscillation of the cantilevers. As expected, driving the system at the anti-symmetric mode leads to a phase shift of $\pi$ between the spin signals of the two NVs.

Furthermore, we can combine the previously presented methods of spin readout and optical driving. Fig\ref{fig:fig2}d shows XY8-5 sequences with varying delays $\phi$ performed on a NV. First, the photo-thermal driving was applied to the cantilever hosting the NV (orange). Afterwards, the driving laser was positioned on the other cantilever. In both cases the mechanical motion is detected by the NV spin. This demonstrates the transduction of a local force, acting on only one cantilever, through the coupled mechanical system. This is a prerequisite for phonon-mediated interactions between distant spins.
Such translation of non-local forces onto the local spin properties allows us to estimate the spin-spin coupling mediated by the common modes, as we detail below.

To estimate the force induced by a single spin on its cantilever motion lets consider the case of a spin-$1/2$ system rotating in the magnetic field gradient produced by the magnetized AFM tip. Due to its rotation in an inhomogeneous magnetic field, the spin will experience an oscillating force with amplitude $F_{z,0} = \mu_B \delta B_z \approx 10^{-18} N$, and ideally, this force is transduced to the cantilever, driving a small amplitude oscillation. Such amplitude could be approximated from the cantilever parameters and the force exerted by the spin on the cantilever as $A_{s} \approx F_{z,0} Q / k = 6.6 \times 10^{-18}$ m. For example, in the above experiment using a double cantilever, where driving one cantilever induces oscillations and spin signal on the other, we can estimate the spin-spin coupling by evaluating the non-local spin-phonon coupling i.e., the spin on one cantilever is coupled to the phonon of the other. Using the experimentally derived local spin-phonon coupling constant $g_{mag}^\prime$, we get a spin-spin coupling $g_{s-s}(Q=10^3) = g_{mag}' A_{s} \approx 3.2$ mHz. It has to be noted, that $A_{s}$ in this case is smaller than the zero-point fluctuation of the cantilever pair. This indicates the need for improvement of the system parameters to achieve observable spin-spin interactions. For example, at low temperatures and ultra-high vacuum, one could achieve an oscillation amplitude of $A_{s} (Q=10^6) \approx 6.64 \times 10^{-15}$ leading to an enhanced spin-spin coupling to $g_{s-s}(Q=10^6) \approx 3.2$ Hz.

In addition to optimizing the system parameters towards a better cantilever, coherent control of the spins can enable noise-free coupling to the mechanical motion and enhance the effective spin-spin coupling. To theoretically analyze these effects we start with the Hamiltonian based on the strain/gradient induced spin-mechanical coupling~\cite{rabl_quantum_2010}, given by 
\begin{equation}
    H= \sum_k H_k,~~H_k = \omega_k a_k^\dagger a_k+ (g_{1,k} S_1^z + g_{2,k} S_2^z)(a_k+a_k^\dagger),
\end{equation}
where $g_i$ are the respective couplings of the spins to the common mode $a_k$. For the two-cantilever problem, $k=2$ with two common modes: symmetric and anti-symmetric. As $[H_i, H_j]=0$, we can treat the dynamics caused by each mode on the spins separately. For simplicity, in the foregoing analysis we consider the mode induced spin-spin coupling for the case of $k=1$. The above Hamiltonian is exactly solvable and the corresponding unitary operator (for $k=1$)  clearly displays both dephasing and entanglement contributions due to the common mode, given by
\begin{equation}
    U(t)={\rm e}^{(-ig_{eff}(t) S_1^z S_2^z )} {\rm e}^{(-i\hat{G}_a(t) (S_1^z +S_2^z))}.
\end{equation}
The first term describes mode induced spin-spin coupling, and the second term describes the spin-phonon coupling dynamics that eventually lead of dephasing of the spins. The spin-spin interaction strength $g_{eff}(t) = \frac{(g^2 (2t\omega-\sin[2t\omega]))}{\omega^2}$ is independent of temperature, while the spin-phonon coupling term that leads to spin-dephasing at a rate $\Gamma(t)=\frac{4g^2 \coth{\beta \omega} \sin 2\omega t }{\omega^2}$, is temperature ($\beta$) dependent. While the large phonon occupancy enhances the spin dephasing rate $\Gamma$, it hardly effects the spin-spin interaction strength $f(t)$. For example using the spin-phonon coupling strength obtained from our experiments, we get the effective spin-spin coupling induced by a single phonon to be $g^{(0)}_{eff} =10~\mu$Hz, and for $N$-phonons (at temperature $T$) collectively contributing to the coupling strength, one could approximate the coupling strength to $g^{(T)}_{eff} =\sqrt{N}g^{(0)}_{eff}$, and for $N \sim 10^6$, this  $g^{(T)}_{eff} \sim 10~m$Hz, which is of the same order as the experimentally obtained value of $20.5~m$Hz. While it appears that the temperature enhances the coupling, it amplifies the dephasing interaction even more rapidly, making the zero-temperature condition more ideal for observing quantum effects in spin mechanics. Due to the linearity of the dephasing term, one could minimize its effects by employing dynamical decoupling, e.g., a CPMG pulse sequence. The effect of such a pulse control on the dephasing term has been analyzed in great detail earlier in~\cite{bennett_measuring_2012}. On the contrary,  we here analyze the dependence of  the induced unitary interaction under the applied pulse control,with $M$ $\pi$-pulses performed at regular intervals $\tau$.  The modified interaction term takes the form
\begin{eqnarray}
g_{eff}(t) &=& \frac{g^2}{\omega^2}\large[M \tau  \omega +(-1)^M \sin (M \tau  \omega )-2 M \tan \left(\frac{\tau  \omega }{2}\right) \nonumber \\
&+&\sec ^2\left(\frac{\tau  \omega }{2}\right) (-\sin (M (\tau  \omega +\pi ))) \large].
\end{eqnarray}
While increasing the number of pulses reduces the dephasing interaction term \cite{uhrig_keeping_2007,rabl_quantum_2010}, it enhances the spin-spin coupling i.e., in the limit of large $M$, while spin-coherence $T_2$ that scales as $\tilde{T}_2 \approx T_2M^{2/3}$ the  two-qubit (spin), effectively scales as  $g^M_{eff} \approx g_{eff} M^{6/5}$. Such interaction induced entanglement among the spins measured through concurrence $(C)$ also improves with the pulse number as shown in Fig. 3a. The relative change of $\gamma$ and $g_{eff}$ being drastically different allows for the emergence of entanglement only above a certain pulse number $M$, that scales linearly with the phonon number $N$. Such prediction to observe coherent spin-phonon effects at room temperature requires extremely large number of $\pi$-pulses is known ~\cite{bennett_measuring_2012}. Under practical considerations the observation of coherent spin-spin effects with initial high-phonon occupancy state requires cooling of the cantilever.  While different methods exist to cool a phonon mode through a two-level system, the well-known side-band cooling in our system could be achieved by switching on the microwave drive at the resonance frequency of the cantilever, thereby locking the spin precesion to the cantilever motion \cite{rabl_cooling_2010}. Under such spin-locking, the dephasing interaction given in Eq. (1) is transformed to a Jaynes Cummings interaction, where the spin and phonon exchange quanta resonantly. Under repeated resetting of the spin polarization, the phonon occupation is drastically reduced. Alternatively an ultra-fast cooling method using selective projective measurements of the spin exists \cite{rao_heralded_2016}. In such cases the cantilever can be drastically cooled as shown in Fig. 3b.

\begin{figure}
    \includegraphics[width=0.5\textwidth]{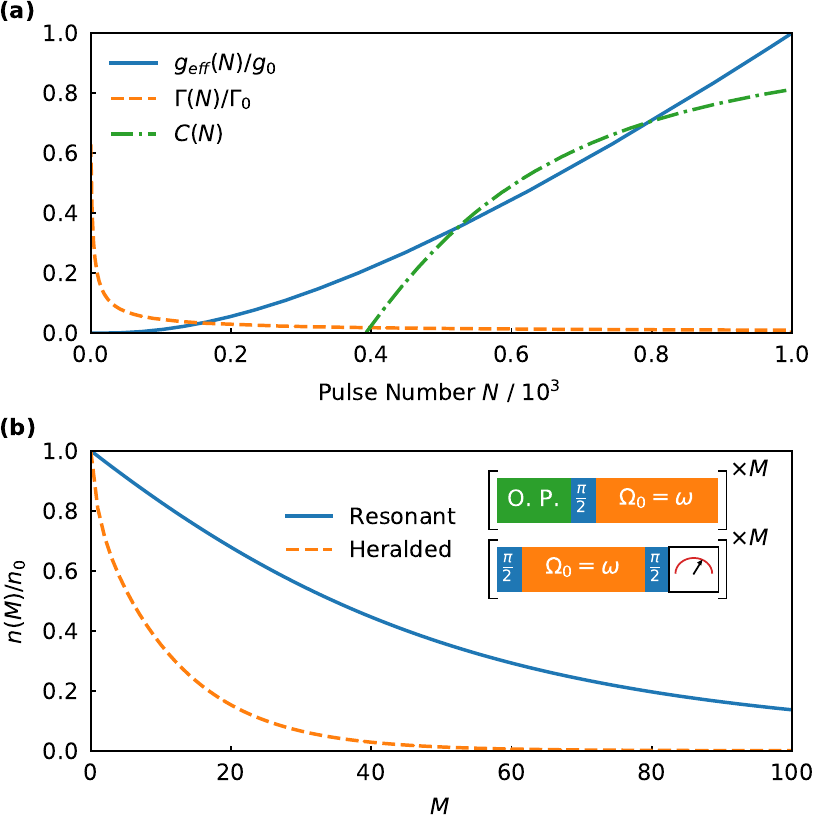}
    \caption{(a) The scaling of the induced spin-spin coupling ($g_{eff}$), the spin decay rate ($\gamma$), and two-qubit entanglement ($C$) as a function of the pulse number $N$. (b) We plot the normalized mode occupation $n \equiv \langle a^\dagger \rangle$ of the cantilever as a function of cooling cycle $M$ for resonant cooling (blue) and heralded cooling (red-dashed). In the inset we show the corresponding pulse sequences employed for the cantilever cooling. }
    \label{fig:fig4}
\end{figure}

In summary we have shown here the prospects for achieving multi-spin, multi-phonon couplings in a spin-mechanical architecture based on diamond cantilevers. For this we have analyzed a coupled cantilever system and shown resonant interaction of the spins of the cantilever modes both through the internal strain and external magnetic field gradient. While gradient coupling results in enhanced (three orders of magnitude) spin-mechanical coupling over the strain-based coupling, the required gradient fields and positioning of the magnetic tips beyond a single coupled-cantilever system needs careful optimization. We have also shown how the coupled cantilevers translates forces applied on one cantilever to the spin in the other cantilever, further leading to weak distant spin-spin interactions ($\sim$mHz). Further, we analyzed theoretically the effects of spin control in enhancing the spin-spin couplings, that lead to their entanglement, and also possible cooling methods for the cantilever through the spin. The strong dephasing of the spin observed under pulsed MW field at the resonant frequency (Fig. 2), when converted to depolarization ($T_{1\rho}$) of the spin under a CW MW field, one would achieve the resonant exchange of quanta between the spin and oscillator, allowing for its cooling. 

We would also like to note that in scaling up such spin-mechanical devices, one has to consider (i) the decreasing zero-point fluctuation, due to increasing mass $m$ of the system, eventually resulting in reduced spin-mechanical coupling $g \propto 1/\sqrt{m} \propto 1/\sqrt{N}$, and (ii) the higher spectral density of mechanical modes with increasing number of coupled cantilevers preventing for the control of individual modes. To avoid this and achieve larger quantum networks one would require an added layer of encapsulation, where multiple small mechanically coupled arrays are interconnected by other means \cite{rabl_quantum_2010,kuzyk_scaling_2018}.

We would like to acknowledge the financial support by the ERC project SMeL, DFG (FOR2724), DFG SFB/TR21, EU ASTERIQS, QIA, Max Planck Society and the Volkswagenstiftung as well as the Baden-Wuerttemberg Foundation.

\bibliography{references}

\end{document}